# A Scalable Framework for Multilevel Streaming Data Analytics using Deep Learning

Shihao Ge
School of Computing
Queen's University
Kingston, ON, Canada
ge.g@queensu.ca

Haruna Isah
School of Computing
Queen's University
Kingston, ON, Canada
isah@cs.queensu.ca

Farhana Zulkernine
School of Computing
Queen's University
Kingston, ON, Canada
farhana@cs.queensu.ca

Shahzad Khan
Gnowit Inc.
Ottawa, ON, Canada
shahzad@gnowit.com

*Abstract*—The rapid growth of data in velocity, volume, value, variety, and veracity has enabled exciting new opportunities and presented big challenges for businesses of all types. Recently, there has been considerable interest in developing systems for processing continuous data streams with the increasing need for real-time analytics for decision support in the business, healthcare, manufacturing, and security. The analytics of streaming data usually relies on the output of offline analytics on static or archived data. However, businesses and organizations like our industry partner Gnowit, strive to provide their customers with real time market information and continuously look for a unified analytics framework that can integrate both streaming and offline analytics in a seamless fashion to extract knowledge from large volumes of hybrid streaming data. We present our study on designing a multilevel streaming text data analytics framework by comparing leading edge scalable open-source, distributed, and in-memory technologies. We demonstrate the functionality of the framework for a use case of multilevel text analytics using deep learning for language understanding and sentiment analysis including data indexing and query processing. Our framework combines Spark streaming for real time text processing, the Long Short Term Memory (LSTM) deep learning model for higher level sentiment analysis, and other tools for SQL-based analytical processing to provide a scalable solution for multilevel streaming text analytics.

*Keywords-deep learning; natural language processing; news media; sentiment analysis; unstructured data*

## I. INTRODUCTION

The world today is generating an inconceivable amount of data every minute [1]. The volume of news data from both mainstream and social media sources is enormous which includes billions of archived documents with millions being added daily. This digital information is available mainly in a semi-structured or unstructured format [2] containing a wealth of information on what is happening around us across the world and our perspectives on these events [3]. Social media platforms like Facebook and Twitter have now become an inseparable part of human communication and a source of a huge amount of data that includes opinions, feelings or general information regarding matters of interest [4]. Some of this data may lose its value or be lost forever if not processed immediately. It is, therefore, important to develop scalable systems that can ingest and analyze unstructured data on a continuous basis [5].

Historically, data acquisition and processing requires several time-consuming steps and traditional analytical processes are limited to using stored and structured data. The time it takes for these steps to complete and drive any actionable decisions are often delayed to the extent that any action taken from the analysis is more reactive than proactive in nature [1]. The extraction of information from streaming text data and tasks involving Natural Language Processing (NLP) such as multilingual document classification, news deduplication, sentiment analysis, and language translation often requires processing millions of documents in a timely manner [2]. These challenges led to streaming analytics, a new programming paradigm designed to facilitate real-time analysis and action on data when an event occurs [6]. In traditional computing, we access stored data to answer evolving and dynamic analytic questions. With stream computing, we generally deploy parallel or distributed applications powered with machine learning models that continuously analyze streams of data [1].

There are typically two phases of streaming analytics involving large scale and high-velocity multimedia data: first, offline data modeling, which involves the analysis of historical data to build models and second, streaming analytics, which involves the application of the trained model on live data streams [1]. For instance, indexing a corpus of documents can be implemented very efficiently with offline processing, but a streaming approach offers the competitive advantage of timeliness [6]. However, media analytics is challenging because of the unstructured and noisy nature of media articles. In addition, many NLP algorithms are compute-intensive and take a long time for execution even for low-latency situations. These challenges call for a paradigm shift in the computing architecture for large scale streaming data processing [2]. Open-source solutions that can process document streams while maintaining high levels of data throughput and a low level of response latency are desirable.

The goal of this study is to develop a distributed framework for scaling-up the analysis of media articles to keep pace with the rate of growth of news streams. The contributions of this work are as follows. First, we propose a scalable framework for multilevel streaming analytics of social media data by leveraging distributed open-source tools and deep learning architectures. Second, we demonstrate the utility of the framework for multilevel analytics using Twitter data.

The paper is organized as follows. Section II presents the research problem with a use case scenario and necessary background information including a literature study on streaming text data analytics. Next, we present our proposed streaming media analytics framework and its features in section III. Section IV provides details about the implementation of the framework while Section V reports our experimental and evaluation results. Finally, we conclude in Section VI with concluding remarks and a list of future work.

## II. BACKGROUND

Offline machine learning has been a valuable analytical technique for decades. The process involves extracting relevant information or intelligence from large stored datasets to predict and prevent future actions or behaviour. However, with increasing access to continuously generated real time data, fast just-in-time analytics can help detecting and preventing crimes, internet frauds, and make the most of market situations in online trading, which require a sub-millisecond reaction time. Critical decisions are made based on knowledge from both past and the new data. Therefore, algorithms are developed to train decision models based on past data, which are then deployed on streaming data to classify situations into different categories such as critical or non-critical or requiring specific actions to enable both proactive and reactive business intelligence [1].

### A. Use Case Scenario

Our industry-academic collaboration with Gnowit, a media monitoring and analytics company is striving to develop an efficient and scalable analytics framework that can facilitate complex multilevel predictive analytics for real-time streaming data from a variety of sources including the Web and social media data. The following are the functional requirements of the analytics pipeline:

- Ingest and integrate streaming news articles data from several sources such as RSS feeds, Twitter Streaming API, and other specialized sources.
- Extract and filter noise such as fake news and duplicates in the data streams.
- Route news articles to relevant consumers such as persistent data stores or analytics engines.
- Create an index for search analytics and visualization.
- Explore and use historical data to develop models for natural language processing and predictive analytics.
- Evaluate and deploy pre-trained models in streaming data analytics pipelines for a variety of real-time text analytics tasks, one of which is sentiment analysis.

There are many applications that can benefit from this analytics pipeline, for example, a company may be looking for the latest trends in its customer's feedback or opinions regarding new products. An example of such application is NewsReader [3], a system for building structured event indexes of large volumes of financial and economic data for decision making. Thus, open-source solutions that can process streaming news articles while maintaining high levels of data throughput and a low level of response latency are desirable.

### B. Challenges in Streaming News Data Analytics

The traditional decision support approaches were based on analysis of static or past data to assess goodness and applicability of derived solutions to evolving and dynamic problems. Recent machine learning approaches train models based on previously collected data and deploy that on stream computing pipelines to analyze ever-changing stream of data [1]. The low-latency, scalability, fault-tolerance, and reliability stipulations of streaming applications introduce new semantics for analytics and also raises new operational challenges [7].

Many traditional machine learning algorithms are iterative and require multiple passes over the data. For instance, stochastic gradient descent optimization algorithm for classification and regression tasks involve repeated scans over the input data to reach convergence [8]. For these algorithms to work in a streaming scenario, they need to be modified for "one pass" processing with compromises in accuracy. Streaming algorithms must also be designed to incorporate strategies for handling temporary peaks in data speed or volume, resource constraints, and concept drifts, a phenomenon that occurs as data evolves and can impact predictive models. Many streaming analytics engines exist today some of which are marketed by large software vendors while others were created as a part of open-source research projects. The most popular generic open source tools which support streaming analytics are Spark Streaming and Flink [6]. These tools enable the development of scalable distributed streaming algorithms and their execution on multiple machines. However, the unstructured and noisy nature of media articles makes media analytics very challenging.

Unstructured data such as news articles from blogs and social media are typically complex and require specialized algorithms that need to be integrated with the streams applications [1]. A fundamental part of this study is to incorporate a scalable real-time sentiment analysis model in the stream analytics pipeline for scoring the opinions and emotions of people towards entities such as individuals, products, services, organizations, issues, events, and topics [9, 10]. Due to the increased access to large text corpora and more computing power [11], deep learning has shown great success in processing language and unstructured text data in general. The integration of deep learning and distributed stream processing will enable fast data analytics tasks to be carried out within a single framework, thereby, avoiding the complexity inherent in using multiple disjoint frameworks and libraries [12].

### C. Related Work

We communicate in natural language through news and social media. News media data allow us to study different perspectives of events happening in the world opening many venues of research [3]. Popular examples of streaming media data sources include tweets from Twitter, web syndication feed via Really Simple Syndication (RSS) or Atom standards, Event Registry[1] and GDELT event stream[2] [13].

---
[1] https://eventregistry.org/
[2] https://www.gdeltproject.org

We discuss some of the research on social media and web data analytics below.

Nair et al. [4] developed a system using Spark' MLlib for extracting data from Twitter's public streams to predict a user's health status. Agerri et al. [2] developed a distributed technology based on Storm to scale-up the modeling of event data in text for reasoning over episodic situations. Karuna et al. [14] proposed a system for extracting heterogeneous information streams from social and web platforms for analyzing real world events in the humanitarian domain along the dimensions of information source, demographics, time, location, and content summaries. These studies utilized traditional machine learning algorithms for developing their models. However, scalable NLP processing requires distributed processing. Given the recent success of deep learning models, these can be incorporated in the analytics pipelines for improving the performances and overall outcomes.

Recent technologies for human language processing are powered by deep learning, a subfield of machine learning, which designs computational models composed of multiple processing layers to learn representations of data using multiple levels of abstraction [11]. Deep learning algorithms are increasingly being used for sentiment analysis. Severyn & Moschitti [15] developed a Convolutional Neural Network (CNN) and achieved a state-of-the-art result on two subtasks of Semeval2015 Twitter Sentiment Analysis (Task 10) challenge.

The Long Short-Term Memory Networks (LSTM) is particularly powerful because of its ability to handle long-term dependencies in sequential data such as news streams. A Tree-Structured LSTM developed by Tai et al. [16] outperformed existing baselines on two tasks: predicting the semantic relation between two sentences (SemEval 2014, Task 1) and sentiment classification (Stanford Sentiment Treebank). Radford et al. [17] developed a deep learning model which learned representations of sentiments in text documents despite being trained only to predict the next character in the text. The authors trained a multiplicative LSTM [18] on a corpus of Amazon reviews to predict the next character in a chunk of text. Radford et al. [19] adapted a study by Dai & Le [20], which demonstrated how to improve the performance of text classification by pre-training and fine-tuning an LSTM model. The study also extends the ULMFiT model [21] that shows how a single dataset-agnostic LSTM language model can be fine-tuned to get state-of-the-art performance on a variety of NLP tasks including sentiment analysis. This study proposes a scalable framework for multilevel streaming analytics of media data by leveraging distributed open-source tools and deep learning architectures.

### III. PROPOSED FRAMEWORK

We propose a distributed framework for multilevel analytics of streaming free-form text data from social and news media containing a wide range of topics and opinions. A multilevel analytics framework in this context refers to the hierarchical processing of data for decision support. Fig. 1 shows the architectural diagram of the proposed framework.

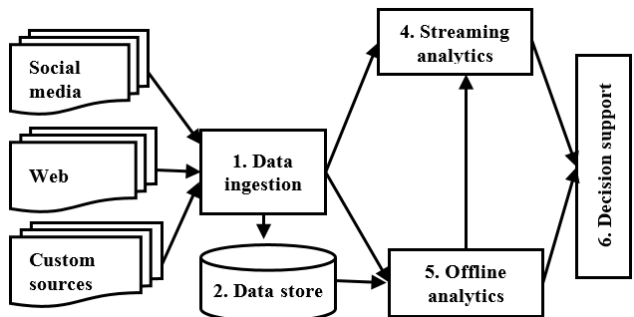

Fig. 1. Architecture of the multilevel streaming analytics framework

The first component of the architecture is the input media generating streaming data which may be in a variety of formats like JavaScript Object Notation (JSON), eXtensible Markup Language (XML), or Comma Separated Values (CSV). At the first level which is the data ingestion stage, data streams from various sources are ingested into a data store or directly into a streaming analytics system. The second level is the data store to archive the ingested data for offline analytics. The third level is the offline analytics stage for large-scale data exploration, query processing, indexing, and development of a language model that can leverage distributed deep learning architectures for NLP tasks such as information search and retrieval, fake news detection, and sentiment analysis. The fourth level is the streaming analytics stage for the real-time scoring of the incoming data streams using models built at the offline analytics stage. The fifth level is the decision support stage which may include real-time notification and visualization.

*A. Data Stream Ingestion*

A variety of open-source ingestion tools, algorithms for building offline models and real-time scoring, distributed streaming engines, and visualization tools exist today, however, these tools are either propriety or lacks the implicit ability to integrate multilevel analytics. Rather than designing a new tool from scratch, we are using open source offerings to develop the proposed framework for multilevel analytics of streaming media data. For ingesting data streams from their sources into analytics platforms or a data store, the popular open-source options include MQTT, RabbitMQ, ActiveMQ, NSQ, ZeroMQ, NiFi, DistributedLog, and Kafka. We are using an integration of NiFi-Kafka dataflow management tools for ingesting data from Twitter. NiFi[3] is a real-time integrated data logistics and simple event processing platform while Kafka[4] is a high-throughput distributed messaging system. NiFi-Kafka integration is an open-source solution that can provide robustness and scalability as well as the ability to add and remove consumers at any time without changing the data ingestion pipeline [5].

*B. Data Store*

In terms of data persistence, the type of data storage depends on the nature of entities of interest and the targeted application. The available open-source options that are suitable for the proposed framework include the Hadoop Distributed File System (HDFS) and Cassandra. We are using HDFS because it was designed for web-oriented applications such as news archiving. HDFS is well suited for distributed storage processing. It is fault tolerant, scalable and extremely simple to expand. HDFS allows data to be stored in a free-schema or raw format.

---
[3] https://nifi.apache.org/
[4] https://kafka.apache.org/

*C. Offline Analytics*

Offline analytics involves the analysis of historical data to build models. There are many algorithms and distributed analytics tools for large-scale offline data analytics. An experimental study on representation learning and deep learning by Dean *et al.* [22] has shown that being able to train large models can dramatically improve performance. As such, we are using a deep learning technique, called Long Short-Term Memory (LSTM) [23], to build a model for computing sentiments in Twitter streams. LSTM is a gated-RNN architecture developed with the capability of learning long-term dependencies in sequential data. It is a powerful technique for solving sequential problems such as the use case described in Section II.

Popular open-source distributed analytics engines include Spark and Flink. This study utilized Spark for building offline analytics models because of its good ability to integrate with other big data tools, popularity, and large developer base. Spark [24] is a fast, in-memory and distributed data processing engine with elegant and expressive development libraries. It allows data workers to efficiently execute streaming, machine learning, and relational workloads that require fast iterative access to datasets. Spark can also integrate with deep learning architectures which are currently gaining increasing attention due to the recent improvements in their performances against previous state-of-the-art results in various NLP tasks [25].

The offline analytics component also supports other higher level analytics such as (i) SQL query processing with Hive, a data warehouse for query and analysis, built on top of the Hadoop ecosystem, and (ii) data indexing and search with Solr, a scalable and fault tolerant platform for distributed indexing.

*D. Online Analytics*

Online analytics involves the use of a deployed model described in the preceding section on incoming data in a lightweight operation such as the scoring and prediction of the sentiment class of incoming news articles. Modern distributed stream processing pipelines receive data streams from several sources, process the streams in parallel on a compute cluster, and then pass the output into continuing data flow pipeline, to a data store, or to a decision support system. Storm, Flink, and Spark Streaming [26] are among the popular open-source libraries that enable the building of scalable and fault-tolerant streaming applications. We are using Spark's Streaming library which allows users to process huge amounts of data using complex algorithms expressed with high-level functions like map, reduce, join, and window. Spark Streaming's minibatch model is ideal for striking a balance in trading off longer latencies for the ability to use more sophisticated algorithms [7].

*E. Decision Support*

Results from both analytics phases can then be visualized or used for decision support and executing appropriate actions in real time. The real-time outputs can empower companies to analyze current and past news contents. We are using Zeppelin and Banana in Solr as decision support tools. Zeppelin is a multi-purpose notebook that integrates smoothly with Spark and supports data analytics and visualization with SQL, Scala, and Python. Zeppelin will be used to visualize offline analytics results and Banana will be used to visualize real-time analytics results from Spark Streaming via Kafka. Fig. 2 shows the framework with the relevant tools considered at each level of analytics as described in Fig.1.

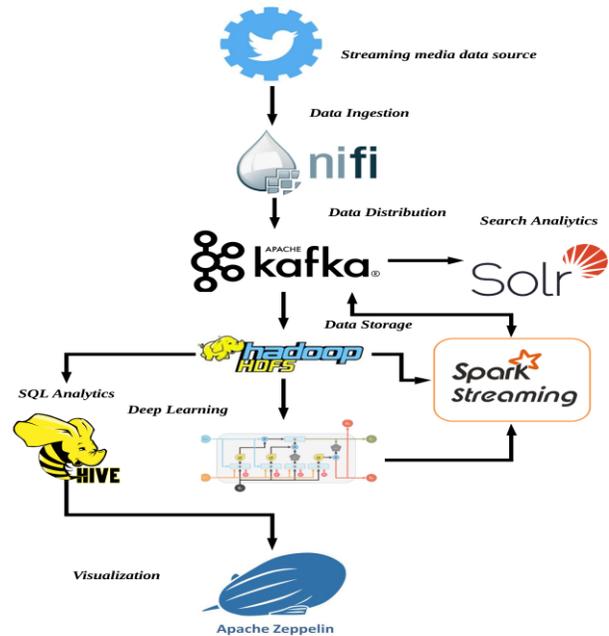

Fig. 2. Implementation of our framework with selected tools.

## IV. IMPLEMENTATION

We implemented a multilevel analytics framework using a set of open-source tools for the use case described in Section II. The framework can also be used as a decision support system for many other use cases. The project implementation environment is a Hadoop Hortonworks Data Platform (HDP) deployed on the SOSCIP Cloud[5], a cloud-accessible cluster of virtual machines having 48GB RAM, 12 VCPU and 120GB memory disk. NiFi-Kafka integration was used for the data acquisition. Keras with TensorFlow backend and a Python implementation of LSTM was used for the sentiment classification model. Scala was used for the SQL queries in Zeppelin and for the Spark Streaming job. Fig. 3 shows the data flow pipeline which includes data ingestion, extraction, archiving, exploration, indexing, modeling, real-time sentiment prediction and visualization of Twitter data.

---

[5] https://cloud.soscip.org

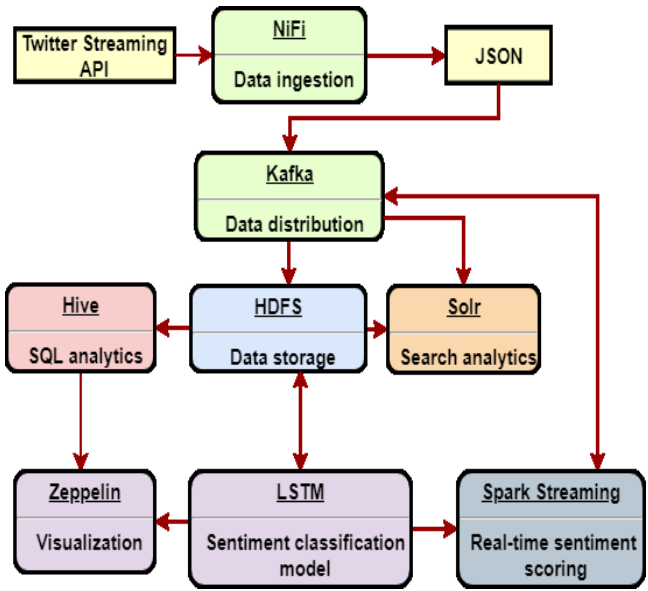

Fig. 3. Flow diagram for the multilevel streaming analytics framework.

As a first step we create a Twitter data ingestion application using Twitter's developer portal and streaming application programming interfaces (API) to authenticate keys and retrieve Twitter data. Next, we execute the following phases:

- *Data stream ingestion:* Using NiFi, we develop a dataflow to ingest data streams from the Twitter Streaming API.
- *Data extraction and distribution:* A JSON parser within NiFi is used to extract the content of the data and then direct the flow into Kafka for data distribution on to HDFS for storage and Solr for search and analytics.
- *Data indexing:* A recent subset of the unstructured data archived from the continuously generated articles from Twitter is then used for indexing and searching in Solr.
- *Data cleaning and feature extraction:* The archived data from the HDFS store is then loaded in Hive, which is followed by data cleaning, exploration, and visualization in Zeppelin using the SparkSQL interpreter.
- *Classifier modeling using deep learning approach:* We developed the sentiment analysis model using the Keras implementation of LSTM. Training a deep learning natural language model can be approached using unsupervised, semi-supervised, or supervised learning. The unsupervised approach requires training using a huge amount of unlabeled dataset equivalent to the entire Wikipedia dump. The supervised approach requires substantial labeled data while the semi-supervised approach typically requires a small amount of labeled data with a large amount of unlabeled data. We followed the supervised approach. The LSTM model for predicting the sentiment of a tweet was implemented using the sentimeny140 dataset. The labeled data is loaded as a Pandas DataFrame onto Keras. The data is then split into 70% training and 30% validation sets. The validation set is used after the model has been trained to evaluate its performance on the data that it has never seen before.
- *Model evaluation:* The model parameters were adjusted to improve its sentiment prediction performance. The model with the best accuracy is then saved on HDFS.
- *Real-time analytics:* The saved model is then utilized for real-time sentiment analysis with Spark Streaming. The streaming application continuously scores and assigns labels to incoming tweets from the NiFi workflow. Kafka is then used to route the labeled tweets to Solr for real-time visualization of the predicted sentiments.

## V. EXPERIMENTAL EVALUATIONS

We train and validate our sentiment classification model using Sentiment140 Twitter dataset which started as a class project from Stanford University[6] and is now available on Kaggle[7]. The datasets contain nearly 1.6 million random tweets labeled as expressing either positive or negative sentiments. Details about the data are shown in TABLE I.

TABLE I. SENTIMENT140 TRAINING DATASET

| Categories | Number | Percentage |
|---|---|---|
| Positive | 788435 | 49.9% |
| Negative | 790177 | 50.1% |
| Total | 1578612 | 100% |

The dataset was uploaded on to HDFS and then loaded in Zeppelin using the SparkSQL interpreter. Fig. 4 is a screenshot of Zeppelin for data exploration, SQL analytics and visualization.

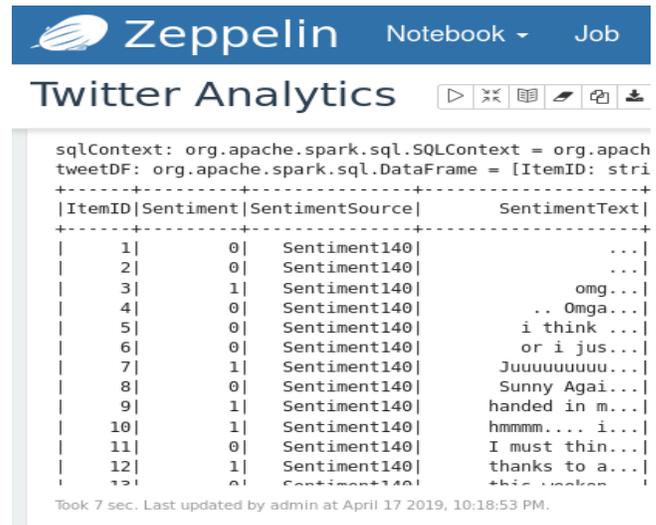

Fig. 4. SQL Analytics with Zeppelin.

Details about the LSTM classification results over the test set are shown in TABLE II. This model is then saved in the HDFS for scoring and predicting the sentiments of real-time tweets.

---

[6] http://help.sentiment140.com/for-students
[7] https://www.kaggle.com/kazanova/sentiment140

TABLE II.    SENTIMENT CLASSIFICATION RESULTS

| Categories | Accuracy |
|---|---|
| Positive | 82.1% |
| Negative | 79.9% |
| Total | 80% |

In this paper, we mainly illustrated the deep learning functionality of our framework and the results of sentiment analysis of Twitter data, which utilized Spark for SQL analytics and LSTM for sentiment classification. A more detailed validation of all the functionality including query processing will be presented elsewhere in an extended version of this work.

VI. CONCLUSIONS

Analytics of streaming data such as news media articles is very challenging because of the unstructured and noisy nature of the media data and requires an integration of news media stream processing applications with specialized machine learning algorithms and big data processing tools. A variety of tools for building offline models and real-time scoring exist today, however, these tools are either propriety or lacks the implicit ability to enable multilevel analytics. This study proposes a scalable framework for multilevel streaming analytics of media data by leveraging distributed open-source tools and deep learning architectures. We demonstrate the functionality of the framework for a multilevel analytics use case scenario using Twitter data. Modeling results that utilized Spark for SQL analytics and LSTM for sentiment classification were presented to demonstrate the usefulness of LSTM for developing models that can learn long term dependencies in language.

Potential future work includes the use of unsupervised approach and attention to develop and fine-tune language models for several downstream NLP tasks. Another interesting area is the integration of deep learning algorithms on scalable machine libraries such as Spark MLlib.

ACKNOWLEDGMENT

We would like to express a special thanks to Southern Ontario Smart Computing for Innovation Platform (SOSCIP) and IBM Canada for supporting this research.  .